\documentclass[journal,comsoc]{IEEEtran}
\usepackage[T1]{fontenc}
\usepackage{times}
\usepackage{soul}
\usepackage{url}
\usepackage[hidelinks]{hyperref}
\usepackage[utf8]{inputenc}
\usepackage[small]{caption}
\usepackage{graphicx}
\usepackage{amsmath}
\usepackage{amsthm}
\usepackage{booktabs}
\usepackage{algorithm}
\usepackage{algorithmic}
\usepackage{multirow}
\usepackage{threeparttable}
\usepackage{amsfonts}
\usepackage{float}
\usepackage{url}
\usepackage{bm}
\usepackage{xcolor}
\usepackage{subcaption}
\usepackage{amssymb}
\usepackage{makecell}
\usepackage{bbding}

\newcommand{\cmmnt}[1]{}

\ifCLASSINFOpdf
\else
\fi

\usepackage{amsmath}

\usepackage[cmintegrals]{newtxmath}

\begin{document}

\title{CosyAudio: Improving Audio Generation with Confidence Scores and Synthetic Captions}
\author{Xinfa Zhu,
        Wenjie Tian,
        Xinsheng Wang,
        Lei He,
        Xi Wang,
        Sheng Zhao,
        Lei Xie~\IEEEmembership{Senior Member,~IEEE}
 \thanks{Corresponding author: Lei Xie}

\thanks{Xinfa Zhu, Wenjie Tian, and Lei Xie are with Audio, Speech and Language Processing Group (ASLP@NPU), the School of Computer Science, Northwestern Polytechnical University, Xi’an 710072, China. Email: xfzhu@mail.nwpu.edu.cn (Xinfa Zhu), twj@mail.nwpu.edu.cn (Wenjie Tian), lxie@nwpu.edu.cn (Lei Xie).}

\thanks{Xinsheng Wang is with Hong Kong University of Science and Technology, Hong Kong, 999077, China. Email: w.xinshawn@gmail.com (Xinsheng Wang).}

\thanks{Lei He, Xi Wang, and Sheng Zhao are with Microsoft Inc., Beijing, 100080, China. Email: helei@microsoft.com (Wendi He), xwang@microsoft.com (Xi Wang), szhao@microsoft.com (Sheng Zhao)}

}

\markboth{Journal of \LaTeX\ Class Files,~Vol.~14, No.~8, August~2015}%
{Wang \MakeLowercase{\textit{et al.}}: Bare Demo of IEEEtran.cls for IEEE Communications Society Journals}


\maketitle

\begin{abstract}

Text-to-Audio (TTA) generation is an emerging area within AI-generated content (AIGC), where audio is created from natural language descriptions. Despite growing interest, developing robust TTA models remains challenging due to the scarcity of well-labeled datasets and the prevalence of noisy or inaccurate captions in large-scale, weakly labeled corpora. To address these challenges, we propose CosyAudio, a novel framework that utilizes confidence scores and synthetic captions to enhance the quality of audio generation.
CosyAudio consists of two core components: AudioCapTeller and an audio generator. AudioCapTeller generates synthetic captions for audio and provides confidence scores to evaluate their accuracy. The audio generator uses these synthetic captions and confidence scores to enable quality-aware audio generation. Additionally, we introduce a self-evolving training strategy that iteratively optimizes CosyAudio across both well-labeled and weakly-labeled datasets. Initially trained with well-labeled data, AudioCapTeller leverages its assessment capabilities on weakly-labeled datasets for high-quality filtering and reinforcement learning, which further improves its performance. The well-trained AudioCapTeller refines corpora by generating new captions and confidence scores, serving for the audio generator training.
Extensive experiments on open-source datasets demonstrate that CosyAudio outperforms existing models in automated audio captioning, generates more faithful audio, and exhibits strong generalization across diverse scenarios.


\end{abstract}


\IEEEpeerreviewmaketitle

\section{Introduction}
\label{sc:Introduction}

\IEEEPARstart{A}{rtificial} intelligence-generated content (AIGC) encompasses a variety of digital media, including images, videos, text, and audio, which have become integral to our daily experiences~\cite{AIGC}. As a key component of AIGC, text-to-audio (TTA) generation offers new opportunities for content creators, advertisers, and video designers~\cite{audiobox}. Initially, research on TTA focused on the label-to-sound setting with a limited set of labels~\cite{label2sound,label2sound2}. With the development of neural networks, TTA models that generate high-fidelity audio from textual descriptions have made significant advancements~\cite{audioldm2,makeanaudio2,Tango2,audiogen1,audiogen2,audiogen3,auffusion}.

Despite these significant advancements, the lack of extensive, high-quality corpora has hindered further development in TTA. Current leading corpora, such as AudioCaps~\cite{audiocaps} and Clotho~\cite{clotho}, comprise dozens of hours of audio data accompanied by meticulously handcrafted captions. The limited size of these corpora poses a challenge for training large-scale TTA models. Constructing large-scale datasets through manual annotation is both time-consuming and labour-intensive, making it a challenging task to accomplish. To overcome this challenge, large-scale weakly-labeled corpora, such as Wavcaps~\cite{WavCaps}, have been introduced. However, due to the lack of effective assessment of caption quality, these weakly-labeled captions often suffer from incompleteness, irrelevance, or temporal inaccuracies, potentially leading to unexpected noise in the generated audio when directly used for training TTA models.

\begin{figure}[]
  \centering
  \includegraphics[width=\linewidth]{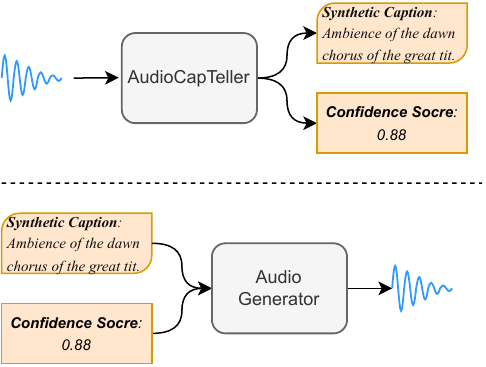}
  \caption{Overview of the proposed CosyAudio. We improve audio generation with confidence scores and synthetic captions, where the AudioCapTeller generates captions and confidence scores from audio and the audio generator synthesizes audio from input captions and confidence scores.}
  \label{fig_overview}
\end{figure}

Drawing inspiration from DALL-E 3~\cite{dalle3}, which improves image generation with better captions, we improve audio generation with better captions. In this context, we define ``better captions" as precise descriptions paired with confidence scores that evaluate their accuracy. Building on this, we introduce CosyAudio, a novel TTA framework that effectively leverages both well-labeled but small-scale corpora and large-scale but weakly-labeled ones. CosyAudio employs a caption refinement and quality awareness approach, leveraging \textbf{co}nfidence scores and \textbf{sy}nthetic captions to enhance \textbf{audio} generation. Specifically, CosyAudio comprises two key components: AudioCapTeller and an audio generator. AudioCapTeller integrates multiple audio-language tasks to achieve robust audio understanding, generating captions alongside confidence scores that assess their accuracy. These outputs enable the audio generator to incorporate synthetic captions and confidence scores for quality-aware audio generation.

Additionally, we propose a self-evolving training strategy to iteratively optimize CosyAudio. This strategy leverages CosyAudio’s caption generation and assessment capabilities to refine itself without relying on external annotated data. The process begins with training AudioCapTeller on well-labeled datasets. Subsequently, AudioCapTeller’s assessment capabilities are applied to weakly-labeled datasets, dividing them into high-quality and low-quality portions. The high-quality portion, combined with well-labeled data, is used for further training, while all portions contribute to fine-tuning AudioCapTeller through direct preference optimization (DPO)~\cite{dpo}. Finally, the well-trained AudioCapTeller refines the entire corpora by generating and evaluating new captions, which are used to train the audio generator and facilitate quality-aware audio generation through confidence scores.

We validate CosyAudio through extensive experiments on open-source datasets, including AudioCaps, Clotho, and WavCaps. The results demonstrate that CosyAudio surpasses state-of-the-art models such as AudioLDM 2~\cite{audioldm2}, Make-An-Audio 2~\cite{makeanaudio2}, Tango~\cite{Tango}, and Tango 2~\cite{Tango2} in generating more faithful audio and achieving superior generalization. Furthermore, our self-evolving training strategy significantly enhances the performance of AudioCapTeller, enabling more accurate caption generation and assessment.
The key contributions of this work are summarized as follows:
\begin{itemize}
    \item We propose CosyAudio, which effectively leverages both well-labeled and weakly-labeled corpora through a caption refinement and quality awareness approach.
    \item We introduce a self-evolving training strategy for text-to-audio generation, which iteratively optimizes CosyAudio through its caption generation and assessment capabilities.
    \item Extensive experiments show CosyAudio achieves robust audio captioning and caption assessment and generates faithful audio with better generalization.
\end{itemize}

The remainder of this paper is organized as follows. Section~\ref{sc:related work} provides a comprehensive review of related work in the field. Section~\ref{sc:method} presents a detailed description of the proposed approach. The experimental setups and results are described in Section~\ref{sc:experiments} and Section~\ref{sc:results}, respectively, where we analyze the performance of CosyAudio and present the evaluation outcomes. Finally, Section~\ref{sc:conclusion} concludes the paper, summarizing the findings and highlighting future research directions. Audio samples can be found on our demo page\footnote{\url{https://zxf-icpc.github.io/CosyAudio/}}.

\section{Related work}
\label{sc:related work}

Automated audio captioning (AAC) and text-to-audio (TTA) synthesis are prominent areas in audio generation research. Additionally, self-evolving training has emerged as an effective strategy for iterative model refinement without reliance on external annotated data. This section reviews prior work in these domains.

\subsection{Automated Audio Captioning}

Automated Audio Captioning (AAC)~\cite{aac} is a multimodal task designed to describe audio content using natural language, capturing acoustic and semantic elements. Unlike tasks that merely identify acoustic events or scenes, AAC seeks to convey the relationships between these events and the overall context, providing a coherent and contextually relevant narrative.

Significant progress in AAC has been driven by the Detection and Classification of Acoustic Scenes and Events (DCASE) challenges~\cite{dcase}. Most AAC models follow an audio encoder and language model decoder architecture, trained on meticulously paired audio-caption datasets. For example, Kim et al.~\cite{audiocaps} use pre-trained PANNs~\cite{panns} to extract audio features, which are then utilized as prefixes for GPT-2~\cite{gpt2} to generate the final captions. EnClap~\cite{enclap} combines EnCodec~\cite{encodec} and CLAP~\cite{CLAP} to extract multi-scale audio features, which are then decoded into text using the pre-trained language model BART~\cite{bart}. 
Beyond foundational architectures, researchers have explored pre- and post-processing techniques to improve performance. For example, RECAP~\cite{recap} employs CLAP to retrieve captions similar to the input audio from a database, constructing prompts for GPT-2 to aid caption generation. LOAE employs pre-trained large language models as post-correctors to address errors in synthetic captions.

Despite their successes on high-quality, human-annotated datasets, these models often struggle to generalize effectively to large-scale, weakly labeled datasets. As these datasets contain richer audio events and more complicated relationships than human-annotated datasets, generating accurate captions for them is a promising direction for the development of AAC.

\subsection{Text-to-Audio}
Text-to-Audio (TTA) aims to transform any descriptive text into vivid audio~\cite{audioldm2}. Unlike Text-to-Speech (TTS)~\cite{survey,metts}, which focuses on converting written text into spoken words, TTA encompasses a broader range of auditory elements. This includes speech, environmental sounds, music, and other audio cues, creating a more immersive and dynamic auditory experience.

TTA has seen substantial progress in recent years. Based on audio representations, mainstream TTA models can be broadly categorized into two approaches: discrete tokens and continuous features. For instance, DiffSound~\cite{diffsound} employs the VQ-VAE model~\cite{vqvae} to convert mel-spectrograms into discrete tokens, utilizing a non-autoregressive token-based diffusion model to generate audio signals. Similarly, AudioGen~\cite{Audiogen} uses an autoregressive model combined with data augmentation to predict discrete audio tokens.
On the other hand, models like AudioLDM~\cite{audioldm}, Make-An-Audio 2~\cite{makeanaudio}, Tango~\cite{Tango}, and Tango 2~\cite{Tango2} rely on Variational Autoencoders (VAEs)~\cite{VAE} to extract continuous representations from mel-spectrograms, leveraging a latent diffusion model (LDM) for audio generation. The key distinction among these LDM-based models lies in their control signals, which guide the generation process. AudioLDM utilizes CLAP~\cite{CLAP}, a joint audio-text representation model, to bridge the gap between text and audio modalities. In contrast, other models demonstrate that pre-trained large language models, such as T5~\cite{T5}, are more effective for controlling audio generation.

Despite the advancements, the generalization and audio quality of current TTA models are still unsatisfactory. The fundamental issue lies in the corpora, specifically the captions corresponding to audio. High-quality corpora with handcrafted captions are limited in size, while large-scale corpora often contain incomplete, irrelevant, or temporally erroneous captions. This discrepancy poses a significant challenge for developing robust and high-fidelity TTA models.

\subsection{Self-evolving Training}
The rapid advancements in large language models (LLMs) have significantly enhanced their reasoning capabilities, leading to increased demand for realistic and generalized reasoning across diverse applications. However, multimodal reasoning, a critical skill for many real-world scenarios, relies heavily on extensive human annotations, which are often scarce and expensive to obtain. To address this limitation, self-evolving training has emerged as a promising approach~\cite{selfevolving}. This method leverages a model’s inherent generative capabilities to iteratively refine and improve itself without requiring external annotated data, making it an appealing strategy for enhancing reasoning abilities~\cite{selfevolving1,selfevolving2,selfevolving3}.

Self-evolving training has gained traction in text and vision domains, where models iteratively generate, assess, and refine their outputs to achieve improved performance. For example, M-STAR~\cite{selfevolving} views self-evolving training as a general reinforcement learning (RL) framework. It leverages the reward model and the prompt variation, to build a clear and unified design space for searching the optimal output.

In the TTA domain, however, most existing models are directly trained on available audio corpora. As highlighted in Section~\ref{sc:Introduction}, large-scale audio corpora often suffer from low-quality captions, including issues such as incompleteness, irrelevance, or temporal inaccuracies. By incorporating self-evolving training, TTA models can address these challenges by refining weakly-labeled data, thereby enabling the effective use of large-scale audio corpora and improving both model performance and generalization capabilities.

\section{Methodology}
\label{sc:method}


In this section, we present the methodology of CosyAudio. First, we provide an overview of the framework, outlining its components and overall design. Next, we delve into the motivation and detailed architecture of each module, beginning with AudioCapTeller, which is responsible for audio captioning and caption assessment. Following this, we describe the proposed audio generator, detailing how confidence scores are incorporated to enable quality-aware audio generation. Finally, we introduce our self-evolving training strategy, which iteratively optimizes CosyAudio using both well-labeled and weakly-labeled corpora

\subsection{Framework Overview}
\label{sc:Overview}

CosyAudio is designed to achieve high-quality audio generation with enhanced generalization and controllability. A critical aspect of this goal is the effective utilization of large-scale corpora. As illustrated in Figure~\ref{fig_overview}, the CosyAudio framework features a core module named AudioCapTeller, which plays a pivotal role in this process. Specifically, AudioCapTeller takes audio clips as input and generates captions along with confidence scores that assess the accuracy of these captions. This capability enables CosyAudio to leverage unlabeled or weakly-labeled corpora by generating high-quality captions and corresponding confidence scores for these datasets.

Once AudioCapTeller has been trained and refined, it refines the quality of large-scale corpora by generating new captions and filtering noisy or low-quality data. Subsequently, the refined corpora are used to train an audio generator, which incorporates confidence scores into the generation process to facilitate quality-aware audio synthesis. This integration ensures that the generated audio aligns closely with textual descriptions while maintaining high fidelity and relevance.

\subsection{AudioCapTeller}
In the CosyAudio framework, AudioCapTeller serves as a pivotal component, bridging the audio and text modalities. Its primary functions include automated audio captioning and generating confidence scores to assess the accuracy of synthetic captions. Drawing inspiration from the Q-Former method~\cite{blip2} for cross-modal tasks, AudioCapTeller employs learnable queries to selectively attend to different types of information, enabling it to handle diverse tasks effectively.

\begin{figure}[]
  \centering
  \includegraphics[width=\linewidth]{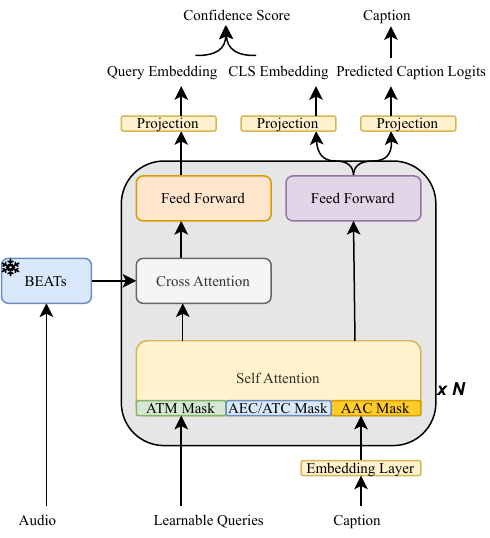}
  \caption{Model structure of AudioCapTeller. We use learnable queries to connect audio and text modalities, enabling the generation of captions and confidence scores from audio.}
  \label{fig_captioner}
\end{figure}

As depicted in Figure~\ref{fig_captioner}, the learnable queries interact with input audio via cross-attention mechanisms. The audio input is represented using features extracted by the pre-trained self-supervised learning model BEATs~\cite{beats}. Additionally, interactions between the learnable queries and text input are facilitated through a shared self-attention block, with attention masks tailored to specific objectives: audio-text matching (ATM), audio-text contrastive learning (ATC), audio events classification (AEC), and automated audio captioning (AAC).

Specifically, ATM assesses the consistency between the input audio and captions by integrating information from both modalities. The attention mechanism is bidirectional, with no masking operation applied. Query embeddings containing multi-modal information are passed through a binary linear classifier to determine whether the audio-text pair is a match. ATC strengthens the global feature alignment between audio and captions. Using an attention mask that isolates queries and text embeddings, ATC calculates pairwise similarities between query embeddings and the global text feature, typically derived from the [CLS] token embedding (as in BERT~\cite{bert}). The objective is to maximize the similarity of positive pairs while minimizing that of negative pairs. AEC enhances AudioCapTeller’s ability to perceive audio events. Using the same masking strategy as ATC to prevent queries from accessing captions, the query embeddings are processed through a semi-supervised multi-label classifier for audio event classification. AAC generates captions for audio clips in an auto-regressive manner. A causal self-attention mask ensures that current token embeddings access historical text embeddings and all audio information derived from queries.

Through these four audio-language tasks, AudioCapTeller establishes robust audio-text alignment. This alignment enables it to generate captions and evaluate their accuracy. In practice, AudioCapTeller synthesizes a caption for a given audio clip and then processes both the audio clip and the generated caption. The pairwise similarity between the query embeddings and the global text feature is calculated, and the highest similarity score is used as the confidence score, reflecting the reliability of the synthetic caption.

\subsection{Quality-aware Audio Generator}

Inspired by the remarkable performance of Tango~\cite{Tango}, an LDM-based TTA model, we design the audio generator in CosyAudio by incorporating confidence scores to enable quality-aware audio generation. These scores allow the generator to evaluate the reliability of input captions, ensuring that the generated audio aligns with textual descriptions.

In the standard Tango model, mel-spectrograms are encoded into latent representations using a variational autoencoder. An Unet-based~\cite{unet} latent diffusion model then models these representations guided by textual input. Specifically, Tango leverages the large language model FLAN-T5~\cite{flan-t5} to extract robust textual representations from captions. These textual representations guide the LDM in constructing target latent audio representations, which are subsequently decoded using a VAE decoder and synthesized into audio using a HiFiGAN vocoder~\cite{hifigan}.

Building upon this structure, we extend by integrating confidence scores. The scores are scalar-quantized and embedded into the audio generation process. Within the U-Net architecture of the LDM, the quantized confidence scores are first processed through a lookup table to generate confidence embeddings. These embeddings are then concatenated with the time-step embeddings, providing additional guidance during the diffusion process. By embedding confidence scores directly into the diffusion mechanism, the audio generator adapts to varying levels of caption reliability, producing higher-quality and more contextually accurate audio outputs.

\begin{figure*}[]
  \centering
  \includegraphics[width=0.8\linewidth]{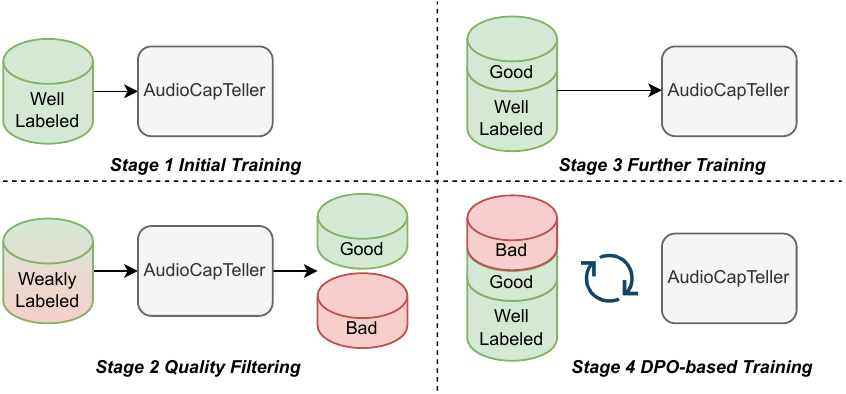}
  \caption{The process of self-evolving training. We use self-evolving training to iteratively optimize AudioCapTeller on well-labeled and weakly-labeled corpora through its caption generation and assessment capabilities}
  \label{fig_training}
\end{figure*}

\subsection{Self-evolving Training}

The corpora used for training TTA models can generally fall into two categories: small-scale datasets with well-labeled captions and large-scale datasets with weakly-labeled captions, where "weakly-labeled" signifies inconsistencies in caption quality. To address the inconsistencies, we propose a self-evolving training strategy to iteratively optimize CosyAudio. This strategy expands the training corpora progressively to enhance both the generalization and accuracy of CosyAudio. Figure~\ref{fig_training} illustrates the overall process.

The self-evolving training strategy for AudioCapTeller can be divided into four stages. In the first stage, AudioCapTeller is trained on the well-labeled datasets to establish foundational audio-language alignment. During this stage, we compute the mean ($\mu$) and standard deviation ($\sigma$) of confidence scores using a validation set. In the second stage, confidence scores are calculated for the weakly-labeled datasets. Audio-text pairs with confidence scores exceeding $\mu - \sigma$ are empirically filtered and regarded as relatively high-quality data. In the third stage, AudioCapTeller is further trained using both the well-labeled and filtered, high-quality datasets. This process strengthens the model's ability to handle diverse audio-caption pairs. The training objective of the first and third stages is as follows:
\begin{equation}
    \mathcal{L}_{cap} = \mathcal{L}_{ATM} + \mathcal{L}_{ATC} + \mathcal{L}_{AEC} + \mathcal{L}_{AAC},
\end{equation}
where $\mathcal{L}_{ATM}$, $\mathcal{L}_{ATC}$, $\mathcal{L}_{AEC}$, and $\mathcal{L}_{AAC}$ represent the loss of audio-text matching, audio-text contrastive learning, audio events classification, and automated audio captioning, respectively. 
The fourth stage addresses the remaining, lower-confidence data excluded in the second stage. Drawing from advancements in reinforcement learning from human feedback, we employ direct preference optimization to fine-tune AudioCapTeller without requiring explicit reward modeling. Specifically, for constructing preference pairs, we generate $N$ captions for each audio clip and rank the synthetic and ground-truth captions according to their confidence scores. The top two captions are considered winners $y_w$, while the bottom two are considered losers $y_l$. Subsequently, the DPO fine-tuning objective is formulated as follows, using these preference pairs to align the model's outputs with higher-confidence captions, thereby refining its performance.
\begin{equation}
\hat{r}_\theta(x, y) =  \frac{\pi_{\theta}(y \mid x)}{\pi_{\text{ref}}(y \mid x)},
\end{equation}
\begin{equation}
\mathcal{L}_{DPO} = - \log \sigma(\beta (\hat{r}_\theta(x, y_w) - \hat{r}_\theta(x, y_l))),
\end{equation}
where $\hat{r}_\theta$ represents the implicit reward model, $x$ represents the input audio. Therefore, the final training objective of AudioCapTeller in the third stage is:
\begin{equation}
    \mathcal{L'}_{cap} = \mathcal{L}_{ATM} + \mathcal{L}_{ATC} + \mathcal{L}_{AEC} + \mathcal{L}_{AAC} + \mathcal{L}_{DPO}.
\end{equation}

The self-evolving training strategy for the audio generator involves caption refinement and quality-aware generation. For caption refinement, AudioCapTeller generates captions for all audio corpora, and confidence scores are calculated for both synthetic and ground-truth captions. For each audio-text pair, the caption with the higher confidence score is selected. Then, for quality-aware generation, the audio generator is trained on this refined dataset. Confidence scores are incorporated into the model to guide the generation process, enhancing the fidelity and faithfulness of the generated audio.

\section{Experimental Setups}
\label{sc:experiments}

This section introduces the experimental setups of CosyAudio, including the database configuration, details of implementation, comparison models, and evaluation metrics.

\subsection{Datasets} 
\label{sc:database}

To evaluate the performance of CosyAudio, we conduct experiments on several open-source corpora, as detailed in Table~\ref{dataSS}. The training and evaluation datasets are chosen to cover a wide range of audio content and annotation quality. The training corpora include AudioCaps, Clotho, and WavCaps. 
AudioCaps is a subset of AudioSet, featuring handcrafted captions for approximately 46,000 ten-second audio clips. Additionally, audio tags for AudioCaps are sourced from the AudioSet metadata.
Clotho consists of 4,981 audio clips, each lasting between 15 and 30 seconds. Each clip is annotated with five reference captions but does not include audio tags.
WavCaps is a large-scale dataset containing audio clips sourced from FreeSound, BBC Sound Effects, SoundBible, and an AudioSet subset. It includes 403,050 audio clips, averaging 68 seconds in duration. Captions in WavCaps are weakly-labeled and generated with the assistance of large language models, alongside accompanying audio tags.

For evaluation, we employ test sets with diverse characteristics. Specifically, we use the MACS~\cite{macs} and Clotho test sets for AAC evaluation. MACS includes 3,930 audio clips from the TAU Urban Acoustic Scenes 2019 dataset, serving as the non-homologous test set. Each clip is annotated by multiple annotators, providing concise, descriptive captions. The Clotho test set is a widely adopted benchmark for AAC and comprises 1,045 audio clips, serving as the homologous test set. Each clip is accompanied by five manually annotated captions. 
For TTA evaluation, we utilize the MACS and AudioCaps test sets instead of Clotho due to their variable audio durations. The test sets include multiple human-written captions for each audio clip. To ensure fairness in the evaluation, we randomly sample 200 captions from each evaluation dataset. This selection balances the diversity of audio types and the quality of captions.

\begin{table}[htb]
\centering
\caption{Audio corpora used to train and evaluate CosyAudio.}
\label{dataSS}
\begin{tabular}{l|ccc}
\toprule
Corpus                        & Duration (hours) & Caption Quality & Usage              \\ \midrule
AudioCaps~\cite{audiocaps} & 136.6  & High & Train \& Eval                          \\
Clotho~\cite{clotho}   & 24.0  & High & Train \& Eval             \\
WavCaps~\cite{WavCaps}  & 7563.3  & Low & Train                           \\
MACS~\cite{macs}       & 10.9  & High & Eval                           \\
\bottomrule
\end{tabular}
\end{table}

\subsection{Implementation Details}
We use a pre-trained BEATs model\footnote{\url{https://github.com/microsoft/unilm/tree/master/beats}} as the frozen audio encoder. The structure of AudioCapTeller is based on the Q-Former configuration\footnote{\url{https://github.com/salesforce/LAVIS/tree/main/projects/blip2}}, with modifications to suit our tasks. Specifically, the semi-supervised audio events classifier comprises three fully connected layers, each with a ReLU activation function.
AudioCapTeller is initialized using pre-trained weights from the BERT-base model~\cite{bert}, while the cross-attention layers and the feed-forward components of the learnable queries are randomly initialized. For constructing preference pairs during the third stage of training, we generate multiple captions per audio clip and ensure a confidence margin of at least $2\sigma$ between the winner and loser captions. During DPO-based finetuning, the parameter $\beta$ is empirically set to 0.05.

The audio generator adopts the architecture and configurations of Tango\footnote{\url{https://github.com/declare-lab/Tango}}, with extensions to incorporate confidence scores. These scores are quantized into five levels based on the statistics of the refined corpora. During inference, the confidence score is set to the highest level to ensure high quality and relevance to input descriptions.

In the aspect of training details, we use the original optimizer from the vanilla Q-Former to train AudioCapTeller. Training spans 10 epochs at each stage on two NVIDIA A100 80GB GPUs, with a batch size of 80 per GPU.  Similarly, we use the original optimizer from Tango to train the audio generator. Training occurs on eight NVIDIA A100 80GB GPUs, with a batch size of 24 per GPU. The audio generator is trained for up to 20 epochs on the refined corpora.

\subsection{Comparison Systems}

To evaluate the performance of CosyAudio, we compare it with several state-of-the-art AAC and TTA models, utilizing their official code and pre-trained checkpoints. 
\begin{itemize}
    \item QwenAudio~\cite{Qwenaduio}: An instruction-following audio language model that scales audio-language pre-training to cover over 30 tasks and facilitate universal audio understanding abilities.
    \item Pengi~\cite{pengi}: An audio language Model that leverages transfer learning by framing all audio tasks as text-generation tasks, which takes an audio clip and instruction as input and generates free-form text.
    \item EnCLAP~\cite{enclap}: An AAC model that utilizes EnCodec and CLAP, along with the pre-trained language model BART, and introduces masked codec modeling to enhance overall acoustic awareness.
    \item RECAP~\cite{recap}: An AAC model that generates captions conditioned on input audio and captions retrieved from a database that are similar to the audio.
    \item AudioLDM 2~\cite{audioldm2}: An audio generation framework proposing a universal audio representation, the Language of Audio (LOA), enabling self-supervised pre-training of the latent diffusion model and translating texts into LOA using a GPT-2 model
    \item Make-An-Audio 2~\cite{makeanaudio2}: A TTA method building on the success of Make-An-Audio, leveraging various techniques to improve semantic alignment and temporal consistency.
    \item Tango~\cite{Tango}: A TTA model that employs FLAN-T5 and latent diffusion model to generate latent audio representations, which are then processed by a VAE and a vocoder to produce audio.
    \item Tango 2~\cite{Tango2}: A TTA system that introduces a preference dataset and fine-tunes the pre-trained Tango using diffusion-based direct preference optimization loss on the preference dataset.
    \item CosyAudio: The proposed framework that utilizes confidence scores and synthetic captions to enhance audio generation, seamlessly enabling both audio captioning and generation.
\end{itemize}

\subsection{Evaluation Metrics} 
\label{sc:evaluation} 

To assess the performance of audio captioning and generation, we utilize a range of mainstream metrics, covering both objective and subjective evaluations.
For audio captioning evaluation, we adopt standard AAC metrics implemented using the \texttt{aac-metrics} library\footnote{\url{https://pypi.org/project/aac-metrics/}}:
\begin{itemize}
    \item BLEU@4~\cite{bleu}: Measures n-gram overlap between generated captions and reference captions, emphasizing syntactic similarity. 
    \item METEOR~\cite{meteor}: Evaluates caption quality by combining precision, recall, and a harmonic mean, incorporating synonym matching for semantic alignment with reference captions.
    \item ROUGE-L~\cite{rouge}: Computes the longest common subsequence (LCS) between generated captions and reference captions, capturing fluency and relevance. 
    \item CIDEr~\cite{cider}: Focuses on syntactic relevance by comparing the consensus between generated captions and human-annotated captions using the term frequency-inverse document frequency weighting. 
    \item SPICE~\cite{Spice}: Emphasizes semantic quality by analyzing scene graphs to capture the underlying meaning of captions.
    \item SPIDEr~\cite{Spider}: A composite metric that combines CIDEr and SPICE to evaluate both syntactic and semantic quality comprehensively.
\end{itemize}

For audio generation evaluation, we employ both objective and subjective metrics. The objective metrics include Frechet Audio Distance (FAD), Kullback-Leibler Divergence (KL), and Inception Score (IS), measured by AudioLDM evaluation tools\footnote{\url{https://github.com/haoheliu/audioldm_eval}}. Subjective evaluation is conducted through two types of Mean Opinion Score (MOS) experiments: OVL and REL, involving 21 volunteers. The rating criteria are: bad = 1, poor = 2, fair = 3, good = 4, great = 5, with 0.5-point increments.
\begin{itemize}
    \item FAD: Assesses the distribution-level similarity between generated and reference audio samples.
    \item KL: A reference-dependent metric that measures divergence between the acoustic event posteriors of ground truth and generated audio.
    \item IS: Evaluates the quality and diversity of generated audio.
    \item OVL: Rates the overall quality of the generated audio.
    \item REL: Assesses the relevance of the generated audio to the input caption.
\end{itemize}

\begin{table*}[htb]
\centering
\caption{Automated audio captioning evaluation on Clotho test set. Note that QwenAudio owns different inference modes, and we test them separately.}
\label{tab_cap1}
\begin{tabular}{@{}l|llllll@{}}
\toprule
Model                                                               & BLEU@4 ↑  & METEOR  ↑ & ROUGE-L ↑ & CIDEr ↑ & SPICE  ↑ & SPIDEr  ↑ \\ \midrule
QwenAudio~\cite{Qwenaduio} (Clotho) & 0.153  & 0.179  & 0.388   & 0.441 & \textbf{0.136} & 0.288  \\
QwenAudio~\cite{Qwenaduio} (plain) & 0.096  & 0.138  & 0.331   & 0.308 & 0.096 & 0.202  \\
Pengi~\cite{pengi}                                                               & 0.150  & 0.172  & 0.375   & 0.416 & 0.126 & 0.271  \\
EnCLAP~\cite{enclap}                                                              & 0.165 & 0.182  & 0.377   & \textbf{0.461} & 0.128 & \textbf{0.294}  \\
RECAP~\cite{recap}                                                               & 0.149  & 0.167  & 0.379   & 0.322 & 0.116 & 0.222  \\
CosyAudio (stage1)   & 0.160  & 0.180  & 0.381   & 0.411 & 0.126 & 0.268  \\
CosyAudio (stage3)  & 0.156  & 0.177  & 0.398   & 0.410 & 0.129 & 0.270  \\ 
CosyAudio (stage4)  & \textbf{0.167}  & \textbf{0.183}  & \textbf{0.411}   & 0.418  & 0.133 & 0.289  \\ \bottomrule
\end{tabular}
\end{table*}

\begin{table*}[htb]
\centering
\caption{Automated audio captioning evaluation on MACS. }
\label{tab_cap2}
\begin{tabular}{@{}l|llllll@{}}
\toprule
Model                                                               & BLEU@4 ↑  & METEOR  ↑ & ROUGE-L ↑ & CIDEr ↑ & SPICE  ↑ & SPIDEr  ↑ \\ \midrule
QwenAudio ~\cite{Qwenaduio} (Clotho) & 0.072  & 0.115  & 0.235   & 0.128 & 0.073 & 0.100  \\
QwenAudio ~\cite{Qwenaduio} (plain) & 0.066  & 0.146  & 0.285   & 0.154 & 0.073 & 0.114  \\
Pengi~\cite{pengi}              & 0.118  & 0.168  & 0.366   & 0.134 & 0.113 & 0.123  \\
EnCLAP~\cite{enclap}             & 0.067  & 0.102  & 0.240   & 0.128 & 0.065 & 0.094  \\
RECAP~\cite{recap}              & 0.063  & 0.098  & 0.224   & 0.119 & 0.062 & 0.092  \\
CosyAudio (stage1) & 0.090  & 0.164  & 0.343   & 0.136 & 0.105 & 0.120  \\
CosyAudio (stage3) & 0.122  & 0.181  & 0.378   & 0.182 & 0.122 & 0.152  \\
CosyAudio (stage4) & \textbf{0.143}  & \textbf{0.195}  & \textbf{0.394}   & \textbf{0.224} & \textbf{0.129} & \textbf{0.177}  \\ \bottomrule
\end{tabular}
\end{table*}

\section{Experimental Results}
\label{sc:results}

This section presents the experimental results of CosyAudio in audio captioning and generation tasks. We further analyze the effectiveness of confidence scores in evaluating caption accuracy and conduct an ablation study to validate the importance of key components, including loss functions, confidence scores, and synthetic captions.

\subsection{Audio Captioning}

Table~\ref{tab_cap1} summarizes the performance of CosyAudio on the homologous test set. CosyAudio surpasses comparison models in BLEU@4, METEOR, and ROUGE-L metrics, demonstrating its ability to generate accurate and contextually rich captions. Moreover, CosyAudio maintains competitive results in CIDEr, SPICE, and SPIDEr, comparable to state-of-the-art models such as EnCLAP and QwenAudio.
Table~\ref{tab_cap1} summarizes the performance of CosyAudio on the homologous test set. The results highlight CosyAudio's superior performance in BLEU@4, METEOR, and ROUGE-L metrics compared to state-of-the-art models. For CIDEr, SPICE, and SPIDEr metrics, CosyAudio achieves results comparable to EnCLAP and QwenAudio. These findings demonstrate CosyAudio's ability to generate accurate, contextually rich audio captions.
Importantly, there is no noticeable performance decline between stage one and stage three, which indicates that the filtering process efficiently isolates high-quality audio-caption pairs from large-scale, weakly-labeled corpora. Furthermore, the performance boost from stage three to stage four underscores the effectiveness of the DPO strategy. By fine-tuning AudioCapTeller, DPO mitigates the generation of low-quality results, aligning outputs more closely with ground-truth captions. The stepwise improvement across all stages emphasizes the importance of confidence scores, which not only aid in filtering inaccurate captions but also guide the construction of preference pairs, driving the system toward higher accuracy and robustness.

To evaluate generalization capabilities, we also test CosyAudio on a non-homologous test set, as shown in Table~\ref{tab_cap2}. CosyAudio significantly outperforms other models, demonstrating superior generalization. Notably, audio large language models like Pengi perform better than traditional AAC models such as EnCLAP, suggesting that training on large-scale, multi-task audio corpora inherently improves generalization.
Within the CosyAudio framework, the progressive improvements across the four stages are particularly striking. Theoretically, small-scale corpora with well-labeled captions play an essential role in the initialization of AudioCapTeller. At the same time, WavCaps, being significantly larger than AudioCaps and Clotho, introduces a broader range of audio events not covered by the smaller datasets. The consistent improvement underscores the effectiveness of the self-evolving training strategy, which enables AudioCapTeller to effectively leverage additional data, improving accuracy and generalization, especially in scenarios with scarce high-quality annotations.

\begin{figure*}[htb]
  \centering
  \begin{minipage}[t]{0.48\linewidth}
    \centering
    \includegraphics[width=\linewidth]{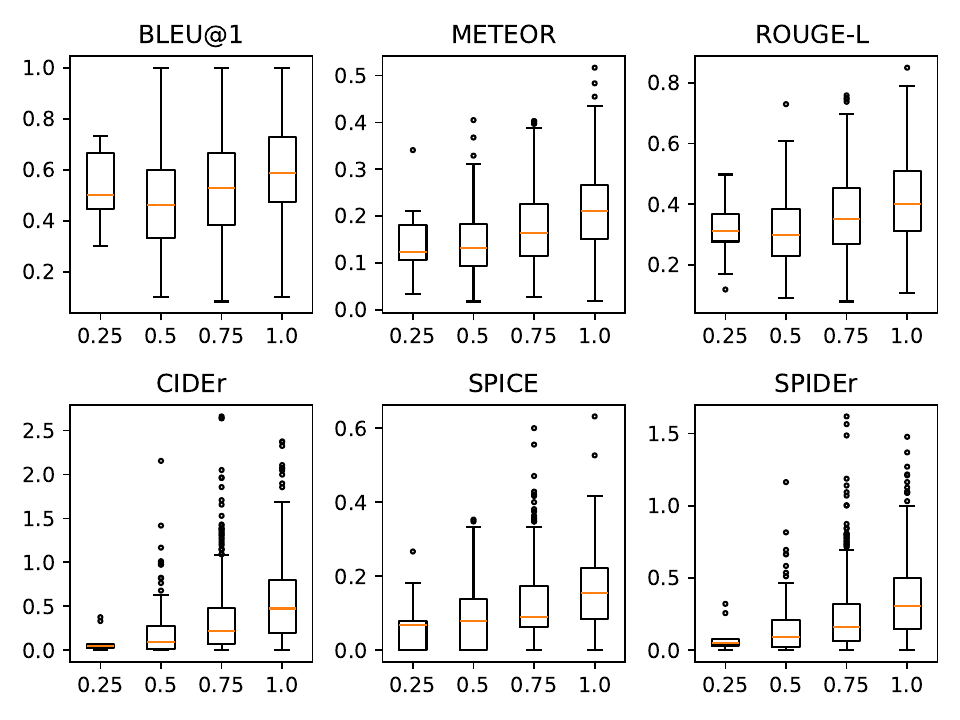}
    \caption*{\small (a) Correlation visualization using CLAP scores}
  \end{minipage}
  \hfill
  \rule{0.5pt}{0.25\textheight} 
  \hfill
  \begin{minipage}[t]{0.48\linewidth}
    \centering
    \includegraphics[width=\linewidth]{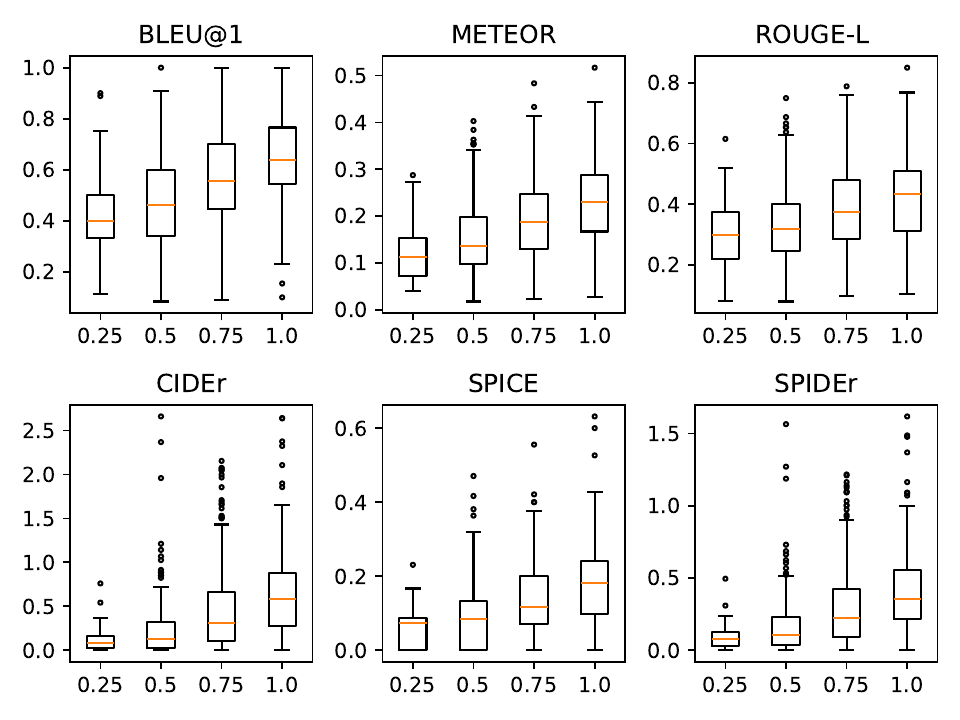}
    \caption*{\small (b) Correlation visualization using confidence scores}
  \end{minipage}
  \caption{Correlation visualization between scores and AAC metrics.}
  \label{fig_eval1}
\end{figure*}

\subsection{Audio Generation}

The results of both objective and subjective evaluations on the homologous test set are presented in Table~\ref{table_tta1}. CosyAudio demonstrates competitive performance against state-of-the-art audio generation models. In objective evaluations, CosyAudio achieves balanced scores across all metrics. While AudioLDM 2 achieves the best FAD score and Tango 2 excels in KL and IS metrics, CosyAudio delivers consistently strong results, showcasing its robustness and reliability.

In subjective evaluations, CosyAudio achieves the highest REL score, confirming the fidelity of the generated audio to the input descriptions. This highlights the pivotal role of accurate synthetic captions and confidence scores in enhancing the quality and relevance of generated audio. Notably, Tango 2 attains the highest OVL score. This can be attributed to the Tango series being exclusively trained and fine-tuned on AudioCaps, whereas CosyAudio leverages both AudioCaps and the large-scale, weakly-labeled WavCaps corpus. Given WavCaps' noise and inherent variability, it is reasonable for CosyAudio to have a slightly lower OVL score on the AudioCaps test set. Nonetheless, CosyAudio’s superior REL score underscores its ability to generalize effectively across diverse audio descriptions, making it suitable and reliable for real-world applications.

\begin{table}[htb]
\centering
\caption{Text to audio evaluation on AudioCaps.}
\label{table_tta1}
\setlength\tabcolsep{5pt}
\begin{tabular}{@{}l|llllll@{}}
\toprule
Model             & FAD↓ & KL↓  & IS ↑ & OVL↑ & REL↑ \\ \midrule
AudioLDM 2~\cite{audioldm2}         & \textbf{3.08} & 1.59 & 5.91 & 3.73$\pm$0.08   & 3.76$\pm$0.10     \\
Make-An-Audio 2~\cite{makeanaudio}     & 3.30 & 1.56 & 5.74 & 3.68$\pm$0.08     & 3.64$\pm$0.11     \\
Tango~\cite{Tango}             & 3.30 & 1.25 & 5.68 & 3.82$\pm$0.07     & 3.79$\pm$0.09     \\
Tango 2~\cite{Tango2}            & 3.74 & \textbf{1.02} & \textbf{6.65} & \textbf{3.95$\pm$0.07}     & 4.02$\pm$0.08     \\
CosyAudio  & 3.56 & 1.30 & 6.22 & 3.91$\pm$0.09     & \textbf{4.07$\pm$0.11}     \\ \bottomrule
\end{tabular}
\end{table}

To evaluate generalization capabilities, additional objective and subjective evaluations were conducted on the MACS corpus in a zero-shot scenario. The results, summarized in Table~\ref{table_tta2}, underscore CosyAudio's significant advantage over other state-of-the-art models. While all models experience performance drops on this challenging zero-shot test set, CosyAudio maintains a clear edge, demonstrating superior adaptability.

Interestingly, while AudioLDM 2 and Make-An-Audio 2 outperform Tango and Tango 2 due to their scalability and ability to leverage extensive datasets, CosyAudio surpasses both by a substantial margin. This performance highlights the efficacy of CosyAudio's caption refinement and quality-awareness mechanisms. By refining synthetic captions and leveraging confidence scores, CosyAudio effectively utilizes large-scale corpora, achieving improved performance in audio generation across diverse and unseen input descriptions.

\begin{table}[htb]
\centering
\caption{Text to audio evaluation on MACS. Note that MACS is included in the training set of Make-An-Audio 2}
\label{table_tta2}
\setlength\tabcolsep{5pt}
\begin{tabular}{@{}l|llllll@{}}
\toprule
Model             & FAD↓  & KL↓  & IS ↑ & OVL↑ & REL↑ \\ \midrule
AudioLDM 2~\cite{audioldm2}         & 7.16 & 1.78 & 3.24 & 3.79$\pm$0.09     & 3.85$\pm$0.08     \\
Make-An-Audio 2~\cite{makeanaudio2}     & 7.51  & 1.72 & 2.97 & 3.70$\pm$0.07     & 3.77$\pm$0.09     \\
Tango~\cite{Tango}             & 9.96 & 1.85 & 2.44 & 3.68$\pm$0.10     & 3.62$\pm$0.10     \\
Tango 2~\cite{Tango2}            & 7.74 & 1.80 & 2.89 & 3.74$\pm$0.09     & 3.70$\pm$0.12     \\
CosyAudio & \textbf{5.74} & \textbf{1.58} & \textbf{4.97} & \textbf{3.88$\pm$0.10}     & \textbf{4.04$\pm$0.11}     \\ \bottomrule
\end{tabular}
\end{table}

\begin{table*}[htb]
\centering
\caption{Ablation study of AudioCapTeller on Clotho.}
\label{tab_ab1}
\begin{tabular}{@{}l|lllllll@{}}
\toprule
Model  & BLEU@4 ↑  & METEOR  ↑ & ROUGE-L ↑ & CIDEr ↑ & SPICE  ↑ & SPIDEr  ↑ & STD ↓ \\ \midrule
CosyAudio  &\textbf{ 0.160 } & \textbf{0.180}  & \textbf{0.381}   & \textbf{0.411} & \textbf{0.126} & \textbf{0.268} & \textbf{0.142} \\
w/o ATM      & 0.138     & 0.172    & 0.358     & 0.368  & 0.117 & 0.242 & 0.161\\
w/o ATC      & 0.142     & 0.172    &  0.363    & 0.375 & 0.122 & 0.248 & 0.224 \\
w/o AEC     & 0.155     & 0.178    & 0.379     & 0.393 & 0.124 & 0.259  & 0.148 \\
w/o AAC      & $5e^{-16}$     & $5e^{-4}$    & $8e^{-4}$     & $1e^{-6}$ & $4e^{-4}$ & $2e^{-4}$ & 0.157 \\ \bottomrule
\end{tabular}
\end{table*}

\subsection{Confidence Scores Analysis}
To assess the effectiveness of confidence scores in evaluating caption quality, we conduct two visualization experiments, using CLAP scores as a benchmark. CLAP~\cite{CLAP}, a widely adopted model for measuring the relevance between audio clips and captions, serves as a comparison model for evaluating CosyAudio's performance.

In the first experiment, we use QwenAudio to generate diverse captions for the Clotho test set and calculate confidence scores for each audio-caption pair. Additionally, we compute AAC metrics (BLEU, METEOR, ROUGE-L, etc.) by comparing the synthetic captions with ground-truth captions. Figure~\ref{fig_eval1} visualizes the correlation between these scores and AAC metrics using normalized box plots within the [0,1] range.

The results reveal that CLAP scores do not consistently align with AAC metrics; for instance, in metrics like BLEU, METEOR, and ROUGE-L, the first box plot often exceeds the second, highlighting inconsistencies. In contrast, CosyAudio's confidence scores exhibit a strictly ascending trend across all metrics, closely mirroring the ground-truth quality of captions. This alignment demonstrates that CosyAudio’s confidence scores more accurately capture caption quality, serving as a reliable indicator of relevance and accuracy.

The second experiment evaluates the consistency of confidence scores for ground-truth captions in the Clotho test set. Since Clotho is meticulously annotated, its captions are expected to exhibit high consistency, reflected in their confidence scores. To analyze this, we normalize the scores to the [0,1] range, compute the standard deviation (STD), and visualize the score distribution using a kernel density plot.

\begin{figure}[htb]
  \centering
  \includegraphics[width=0.8\linewidth]{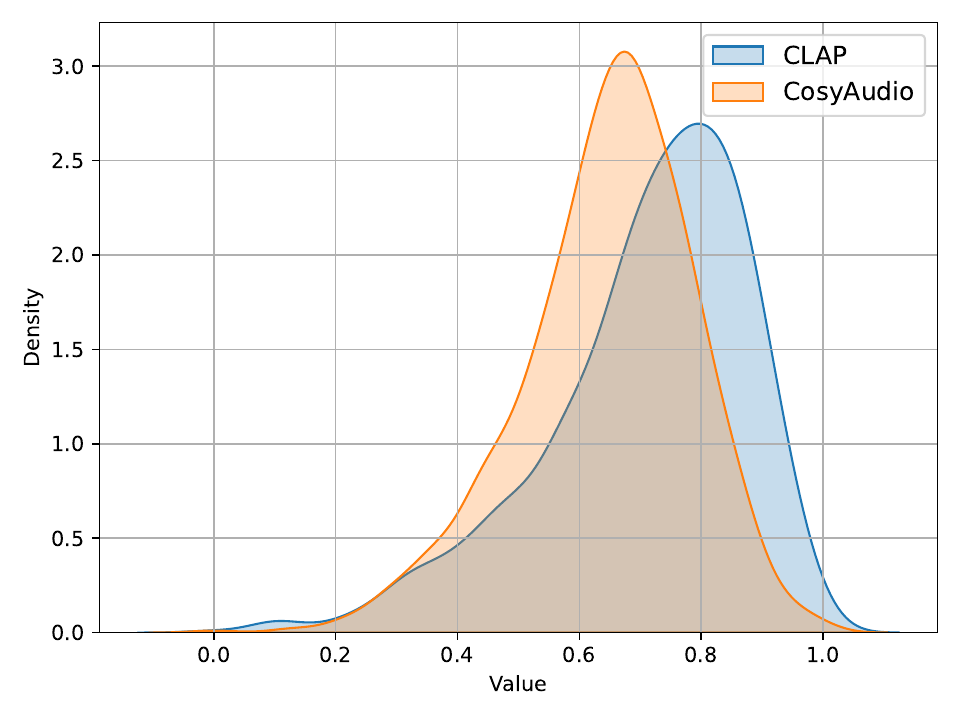}
  \caption{Confidence scores distribution of different evaluation models.}
  \label{fig_eval2}
\end{figure}

As illustrated in Figure~\ref{fig_eval2}, the density curve for CosyAudio is sharper and more concentrated, with an STD of 0.142, compared to CLAP's broader curve and higher STD of 0.167. This sharper curve indicates that CosyAudio provides more stable and consistent confidence scores, aligning closely with human annotations. Such consistency underscores the robustness of CosyAudio’s evaluation mechanism, even in datasets with stringent annotation standards like Clotho.

These visualization experiments validate the robustness of CosyAudio's confidence scores in assessing the quality of both generated and ground-truth captions. By aligning closely with AAC metrics and maintaining high consistency in meticulously annotated datasets, CosyAudio is capable of not only enhancing the reliability of audio-caption relevance assessment but also facilitating the refinement of audio captions in our CosyAudio framework.

\subsection{Ablation Study}

We conducted an ablation study to systematically assess the contributions of key components within CosyAudio. This section investigates the impact of various audio-language tasks in AudioCapTeller and evaluates the role of confidence scores and synthetic captions in the audio generator.

We first analyze the effects of various audio-language tasks by selectively removing loss components during the training of AudioCapTeller in stage one. The results, presented in Table~\ref{tab_ab1}, reveal the following insights. First, removing the automated audio captioning loss causes a significant decline across all AAC metrics. This demonstrates the foundational role of AAC loss in enabling AudioCapTeller to generate accurate and contextually appropriate captions. Besides, excluding the audio-text contrastive and audio-text matching losses increased the standard deviation of confidence scores, indicating greater inconsistency in evaluating the quality of ground-truth captions. This highlights the importance of audio-text contrastive learning and matching for aligning audio and text modalities, thereby ensuring reliable and consistent confidence scores. Finally, removing the audio event classification loss resulted in a slight reduction in AAC metrics. This suggests that classifying audio events provides valuable context in scenarios involving overlapping audio events, enhancing the depth and accuracy of generated captions.

We further analyze the influence of confidence scores and synthetic captions on the audio generator. The results are detailed in Table~\ref{tab_ab2}. First, we generate audio using identical captions but with various confidence scores. The results reveal that higher confidence scores lead to superior performance across FAD, KL, and IS metrics. Audio synthesized using low-confidence scores exhibits artifacts such as noise and missed audio events, whereas high-confidence scores produce audio that is clear, accurate, and faithful to input captions. These results reveal that higher confidence scores consistently contribute to audio generation. In addition, excluding confidence scores during audio generator training results in average performance. This shows that confidence scores are essential for distinguishing high-quality training samples and maintaining control over synthetic audio quality. Without them, the model lacks prioritization, leading to less precise outputs. Moreover, training the audio generator directly on large-scale corpora without confidence scores or synthetic captions yields the worst results. This outcome emphasizes the necessity of both components for effective audio generation. Confidence scores enhance sample quality control, while synthetic captions refine training data, together enabling robust and high-quality audio synthesis.

\begin{table}[htb]
\centering
\caption{Ablation study of the audio generator on MACS.}
\label{tab_ab2}
\begin{tabular}{@{}l|llllll@{}}
\toprule
Model             & FAD↓ & KL↓ & IS ↑ \\ \midrule
w/ High confidence   & \textbf{5.74}     & \textbf{1.58}    & \textbf{4.97}     \\
w/ Middle confidence &  6.67    & 1.72    & 3.74     \\
w/ Low confidence    & 8.52     & 1.87    & 2.12     \\ \midrule
w/o Confidence          & 6.83     & 1.71    & 3.55     \\ 
w/o Cosy          & 7.19     & 1.76    & 2.32     \\ \bottomrule
\end{tabular}
\end{table}

\section{Conclusion}
\label{sc:conclusion}

In this paper, we present CosyAudio, a novel framework designed to leverage both well-labeled and weakly-labeled corpora through a caption refinement and quality awareness approach. Central to our framework is AudioCapTeller, a versatile module that enables both audio captioning and caption assessment. Building on AudioCapTeller, we propose an audio generator that enhances audio generation by incorporating synthetic captions and confidence scores. Furthermore, a self-evolving training strategy is introduced, comprising a four-stage training pipeline for AudioCapTeller and utilizing refined corpora to optimize audio generator training.
Comprehensive experiments on open-source datasets demonstrate the effectiveness of CosyAudio. The framework generates precise captions, accurately assesses caption quality, produces faithful audio, and exhibits strong generalization capabilities across diverse test sets. 

Despite its strengths, we identified a key area for further improvement: the quality of synthetic audio. While CosyAudio effectively mitigates noise in captions, such as irrelevance, incompleteness, and temporal inaccuracies, residual noise in ground-truth audio poses a persistent challenge. Addressing this limitation could further enhance the performance of CosyAudio. We believe that using a higher audio quality corpus can improve the fidelity of synthetic audio, paving the way for more robust and versatile audio generation systems.

\bibliographystyle{IEEEtran}
\bibliography{mybibfile.bib}

\end{document}